\documentclass[12pt]{iopart}

\usepackage{graphicx} 
\usepackage{bm}
\usepackage{xspace} 
\usepackage{float}

\newcommand{\um}{$\mathrm{\mu m}$\xspace}
\newcommand{\rtHz}{$/ \sqrt{\rm Hz}$\xspace}
\newcommand{\aNrtHz}{${\rm aN} / \sqrt{\rm Hz}$\xspace}
\newcommand{\fmrtHz}{${\rm fm} / \sqrt{\rm Hz}$\xspace}
\newcommand{\Sx}{$S_{zz}$\xspace}

\newcommand{\Sfth}{$S_{FF}^{\rm th}$\xspace}
\newcommand{\Sv}{$S_{VV}$\xspace}

\begin{document}

\title[Multidimensional optomechanical cantilevers for HF-AFM]{Multidimensional optomechanical cantilevers for high frequency atomic force microscopy} 

\author{C Doolin, P H Kim, B D Hauer, A J R MacDonald and J P Davis}
\address{Department of Physics, University of Alberta, Edmonton, Alberta, Canada T6G 2E9}
\ead{jdavis@ualberta.ca}

\begin{abstract}

High-frequency atomic force microscopy has enabled extraordinary new science through large bandwidth, high speed measurements of atomic and molecular structures.  However, traditional optical detection schemes restrict the dimensions, and therefore the frequency, of the cantilever - ultimately setting a limit to the time resolution of experiments. Here we demonstrate optomechanical detection of low-mass, high-frequency nanomechanical cantilevers (up to 20 MHz) that surpass these limits, anticipating their use for single-molecule force measurements.  These cantilevers achieve 2 fm\rtHz displacement noise floors, and force sensitivity down to 132 aN\rtHz.  Furthermore, the ability to resolve both in-plane and out-of-plane motion of our cantilevers opens the door for ultrasensitive multidimensional force spectroscopy, and optomechanical interactions, such as tuning of the cantilever frequency \textit{in situ}, provide new opportunities in high-speed, high-resolution experiments.

\end{abstract}

\section{Introduction}

Since the atomic force microscope (AFM) was first demonstrated \cite{Binnig1986}, it has become an indispensable tool for probing the physical characteristics of microscopic systems.   Working by Hooke's Law, $ \mathbf{F}$\,=\,$- k\mathbf{x} $,  
the tip of the AFM is displaced proportional to an applied force, transducing forces into a detectable signal.  This has been used to great effect for surface imaging, where interatomic forces between an AFM tip and substrate are measured as raster images of the surface structures down to the atomic scale \cite{Giessibl1995} and beyond \cite{Gross2009}.  The ability to use AFMs in liquid environments \cite{Fukuma2005} has led to their widespread use in biological applications \cite{Sharma13}, such as live imaging of biological specimens \cite{Muller2008}, and non-scanning applications like studying receptor-ligand binding of surface proteins \cite{Viani2000} and deciphering the mechanics of proteins through unfolding experiments \cite{Rief1997, Bro03}.  For both aqueous and high-speed AFM (HS-AFM), it is advantageous to use low-mass, high-frequency cantilevers \cite{Eki05}, yet current technology is limited in detecting the motion of such cantilevers \cite{Ando12}.

In biological applications, where the sample is often continuously moving,  the time resolution of the measurement process is critical.  High-speed AFM \cite{Ando2001} has enabled the dynamics of molecular systems to be visualized at speeds of up to 80 ms for a $50 \times 100$ pixel image \cite{Uchihashi2011}.  This has permitted the real-time imaging of individual motor proteins \cite{Uchihashi2011}, proteins diffusing and interacting in lipid bilayers \cite{Casuso2010}, and the folding of synthetic DNA origami structures \cite{Endo2009}.  When operated dynamically \cite{Albrecht1991},  the maximum time resolution of the measurement is limited by the frequencies of the structural modes of the cantilever.  In the simple harmonic approximation these are $\omega_0 = \sqrt{ {k_{\rm eff}} / {m_{\rm eff}} } $, where $k_{\rm eff}$ and $m_{\rm eff}$ are the effective spring constant and mass of a particular mode \cite{Hau13}.  Since $k_{\rm eff}$ is chosen to optimize the displacement response of the AFM to the application's characteristic forces,  minimizing the dimensions, and therefore $m_{\rm eff}$, grants access to the regime of both delicate force sensing and exceptional time resolution through increased mechanical frequencies.  
\begin{figure}[h]
\centerline{\includegraphics[width=4.5 in]{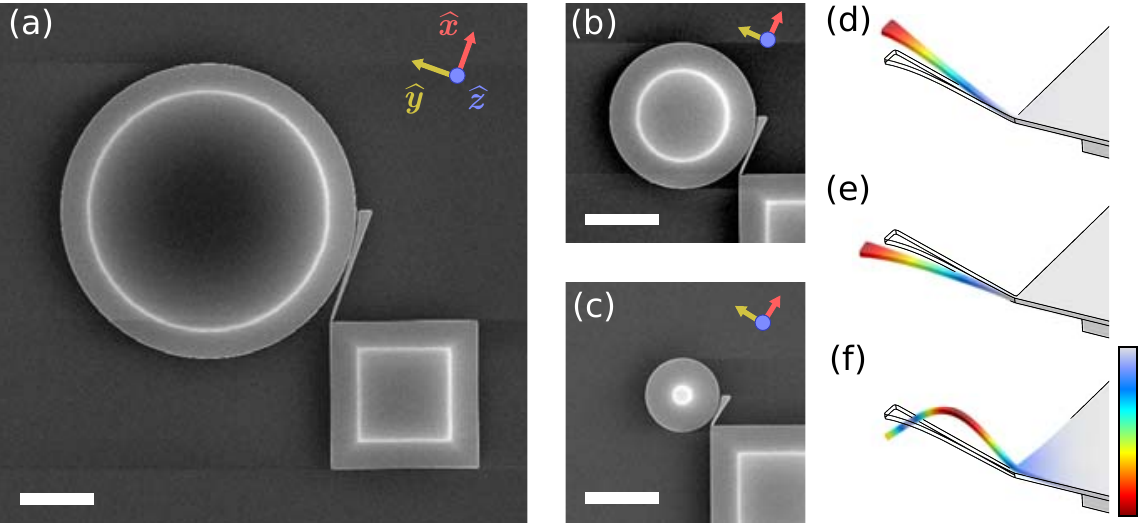}}
\caption{{\label{fig.devs}} (a)  SEM image of the optomechanical device with a 20 \um diameter optical microdisk evanescently coupled to an 8 \um long cantilever.  Coordinates are aligned such that $\hat{x}$ is parallel to the axis of the cantilever, $\hat{y}$ points along in-plane motion of the cantilever, and $\hat{z}$ points out-of-plane.  (b)  10 \um disk, 4 \um cantilever and (c) 5 \um disk, 2 \um cantilever; scale bars 5 \um on all panels. (d)-(f)  FEM simulations reveal the first three modes of the 8 \um long cantilever as an example: an out-of-plane mode, an in-plane mode and a second out-of-plane mode.  Mechanical modes of the shorter cantilevers are similar.  Colour scale indicates relative displacement.
}
\end{figure}

\begin{figure*}[t]
\centerline{\includegraphics[width=5.5in]{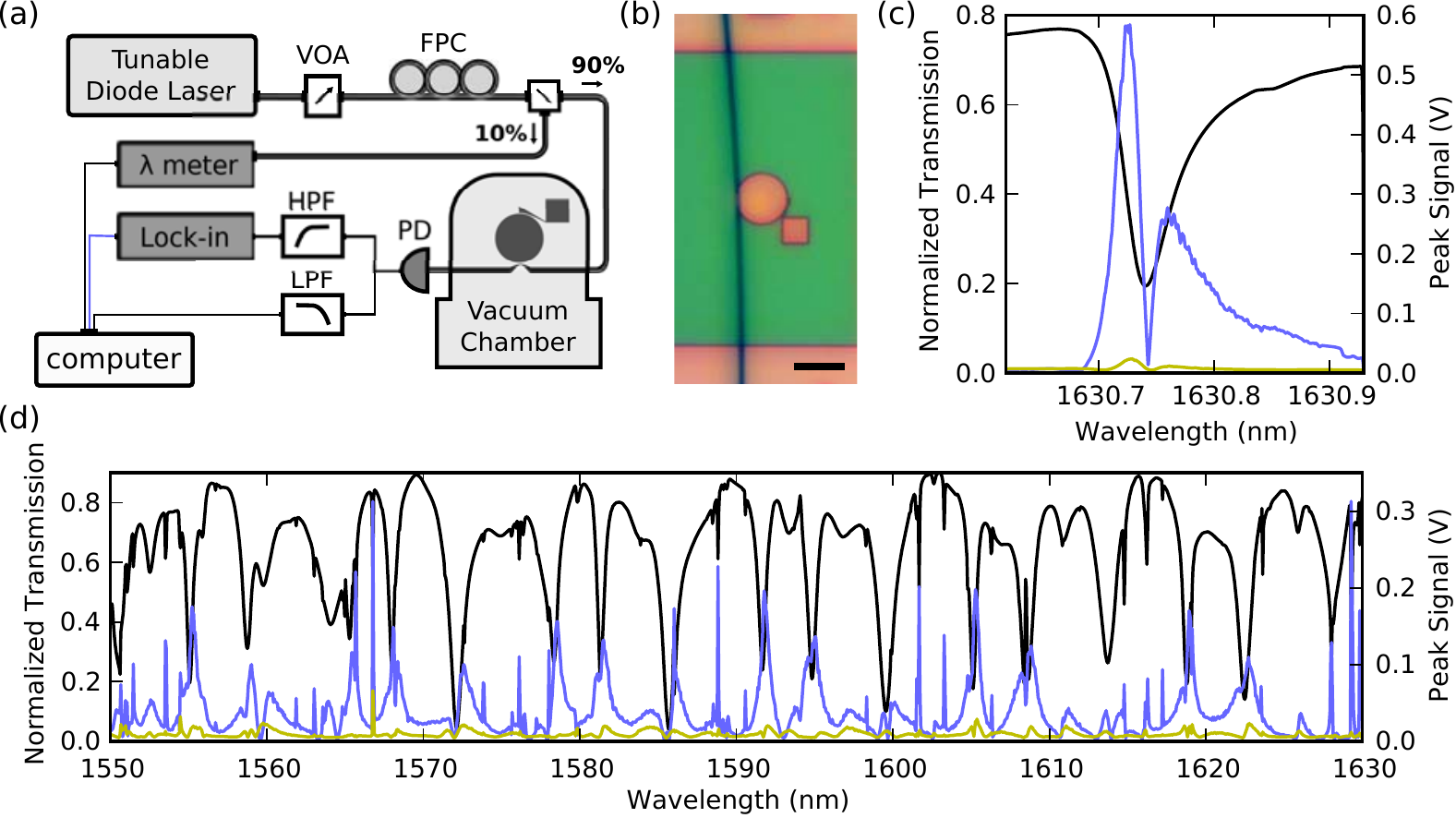}}
\caption{{\label{fig.how}} (a) Schematic of experimental setup (VOA - variable optical attenuator, FPC - fiber polarization controller, HPF - high pass filter, LPF - low pass filter, PD - photodiode). (b)  Optical image of a dimpled, tapered fiber placed on a 10 ${\rm \mu m}$ diameter disk opposite an 8 \um long cantilever; scale bar 20 \um.  (c)  Transmission (normalized to transmission in the absence of coupling) through the tapered fiber (black), while simultaneously locked-on to the out-of-plane mode (light blue) and the in-plane mode (yellow), reveals the maximum peak signal occurs slightly detuned from the optical resonance, approximately corresponding to the maximum slope of the transmission \cite{Kim13}.  (d) Scanning the entire frequency range of the tunable laser reveals optical resonances that provide maximum signal.
}
\end{figure*}

Common methods to detect the displacement of a cantilever include reflecting a laser beam off the cantilever onto a position sensitive photodetector, termed optical beam deflection (OBD), or recombining the reflected beam interferometrically.
An important benchmark of a displacement detection system is the displacement noise floor: the noise corresponding to the minimum displacement resolvable by the detection system.  OBD has obtained displacement noise floors of 5 fm\rtHz \cite{Fukuma2009}, while an all-fiber interferometer has achieved noise floors of 2 fm\rtHz \cite{Rasool2010}, both with standard low-frequency cantilevers ($\sim$ 300 kHz).  However, these detection methods scale poorly as the dimensions of the nanomechanical devices fall below the spot size of the laser beam ($\ge$ 1 $\mu {\rm m}$) \cite{Eki05}, creating an effective limit on cantilever sizes (and frequencies) that has already been reached \cite{Ando12}.

The technique of optomechanics \cite{Eichenfield2009, Anetsberger2009, San10, Bag11, Bar12, Sau12} offers unprecedented displacement sensitivity while being well suited for nanoscale devices.  By spatially localizing optical cavity modes with a mechanical resonator, motional degrees of freedom are coupled to frequency (or phase \cite{Anetsberger2010}) shifts of the optical modes. Monitoring the transmission of laser light coupled to the optical cavity then provides sensitive readout of the mechanical motion,  exemplified by experiments measuring the motion of nanomechanical resonators to the standard quantum limit (SQL)---the theoretical noise floor of a continuous measurement determined from dynamical backaction and photodetector shot noise  \cite{Anetsberger2010}---as well as observing quantum behavior \cite{Saf12}.

\begin{figure*}[t]
\centerline{\includegraphics[width=6.0in]{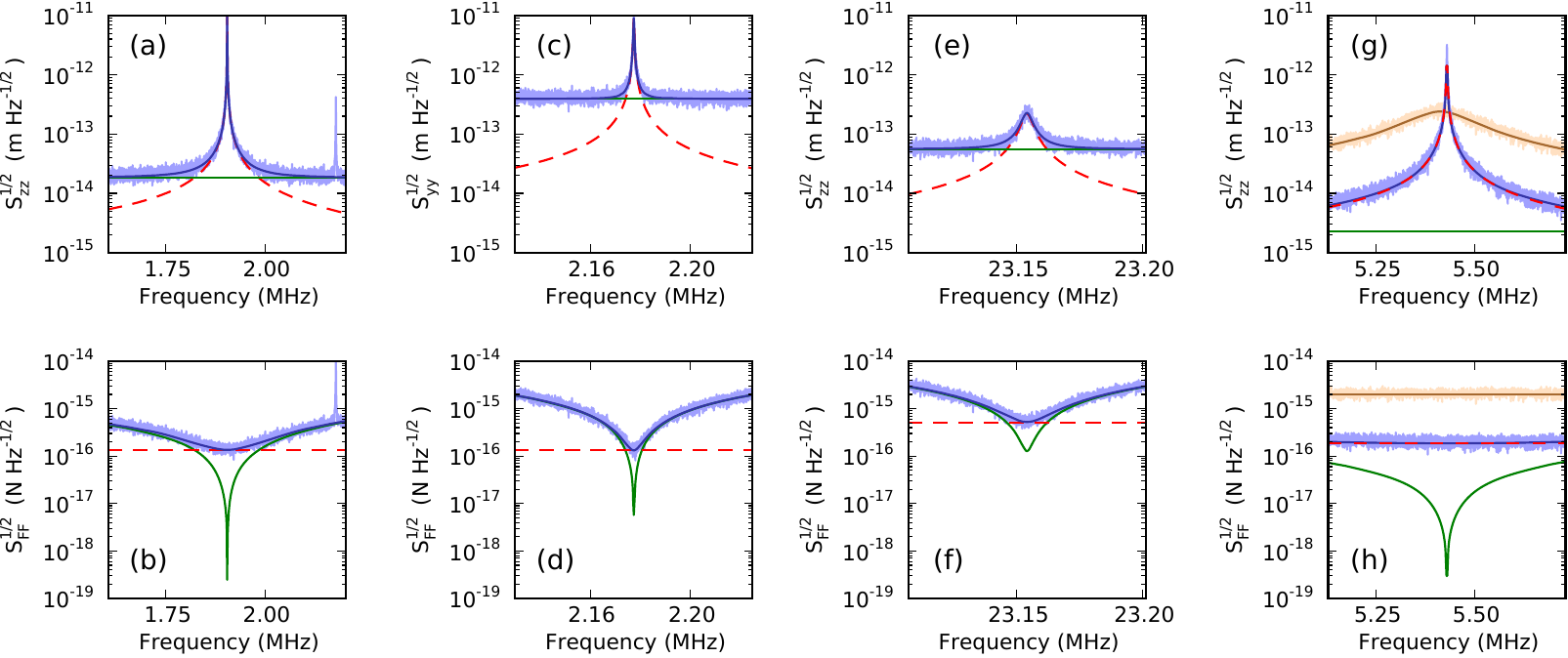}}
\caption{{\label{fig.peaks}} (a)  A peak in the displacement noise density, $\sqrt{S_{zz}}$, corresponding to out-of-plane motion of the 8 \um cantilever.  The peak at higher frequency is the in-plane mode.  $\sqrt{S_{zz}}$ is fit to a superposition (blue) of the thermal noise of the cantilever ($\sqrt{S_{zz}^{\rm th}}$, red dashed) and a constant measurement noise, the displacement noise floor ($\sqrt{S_{zz}^{\rm nf}}$, green).  (b) By dividing $\sqrt{S_{zz}}$ by the force susceptibility, $| \chi(\omega)|$, the measured force noise density, $\sqrt{S_{FF}}$, can be obtained.  (c) $\sqrt{S_{yy}}$ and (d) $\sqrt{S_{FF}}$, corresponding to the 8 \um cantilever's in-plane plane mode,  and (e), (f) second out-of-plane mode.  In all cases $\sqrt{S_{FF}}$ is limited by thermal forces when at the cantilever resonance frequency and limited by detector noise off-resonance.
(g)  $\sqrt{S_{zz}}$, and (h)  $\sqrt{S_{FF}}$ of the 4 \um device's out-of-plane mode are dominated by thermal noise across a wide frequency range due to the low optomechanical detection noise floor.  Shown in light brown are $\sqrt{S_{zz}}$ and $\sqrt{S_{FF}}$ in air, with corresponding fits in dark brown.  $\sqrt{S_{zz}^{\rm nf}}$ in air agrees with that in vacuum, but $\sqrt{S_{FF}^{\rm th}}$ is limited to 2 fN\rtHz due to the viscous damping, compared with 180 \aNrtHz.  Further data is shown in Appendix D.}
\end{figure*}

\section{Results and discussion}

Here, three sizes of low mass, MHz frequency, optomechanical devices suited to AFM applications are presented.  They consist of cantilever-style nanomechanical resonators coupled to the whispering gallery modes of optical microdisks and are commercially fabricated from a 215 nm thick silicon layer of a silicon-on-insulator (SOI) wafer, ensuring simple fabrication with automatic and reproducible optomechanical cavity formation.  The cantilevers have lengths of  8, 4, and 2 \um, and are on average 400 nm wide, broadening towards the end to allow binding of molecules to the cantilever, for pulling experiments, without compromising the optical cavity quality ($Q_{\rm opt}$\,$\sim$\,3\,$\times$\,$10^4$ for 20 \um diameter disk). They couple to disks of 20, 10 and 5 \um diameter respectively, and scanning electron microscopy (SEM) images, and finite element method (FEM) simulations of the first three structural modes of the 8 \um long cantilever, are shown in figure \ref{fig.devs}.  We envision single molecule force (folding/unfolding) experiments as the ideal AFM application for these devices, as this would not degrade the optical $Q$ of micro disk due to a sample, nor would a separate tip need to be attached.  

\begin{table*}[t]
\begin{tabular}{ l @{\hskip 3pt} l  *{7}{c} }
\multicolumn{2}{l}{Cantilever Length} & $m_{\rm eff}$ & $k_{\rm eff}$ & $\omega_0 / 2\pi$ & $Q$ (air) & $\sqrt{S_{zz,yy}^{\rm nf}}$ (air) & $\sqrt{S_{FF}^{\rm th}}$  (air) & $g_{\rm om}$ \\
\multicolumn{2}{c}{$[\mathrm{\mu} {\rm m}]$} & $[{\rm fg}]$ & $[$N/m$]$ & $[$MHz$]$ &  & $[$\fmrtHz$]$ & $[$\aNrtHz$]$ & $[$MHz / nm$]$ \\
\hline
\hline
2 & out-of-plane  & 140 & 2.2     & 20.1 & 3,600 (120) &    20 (18)  &       290 (1,500)      & 2,000         \\
2  &  in-plane & 180 & 3.3     & 21.4 & 5,000           &   120  &           280          & 340            \\
4  & out-of-plane & 240 & 0.30   & 5.43 & 4,300 (35)   &      2 (3)    &        180 (2,000)     & 720   \\
4  &  in-plane & 260 & 0.48   & 7.04 & 4,400          &    300    &           200          & 6              \\
8  &  out-of-plane & 610 & 0.087 & 1.90 & 6,500 (22)  &      18 (17)   &       135  (2,300)     & 150            \\
8  &  in-plane & 610 & 0.11   & 2.18 & 7,800          &    390    &            132         & 7                 \\
8  &  2$^{\rm nd}$ out-of-plane & 610 & 13      & 23.2 &      5,600     &    55     &              510        & 57                \\

\end{tabular}
\caption{{\label{thetable}} Measured parameters of investigated devices. Data is presented for three optomechanical devices of varying size, but similar geometry (figure \ref{fig.devs}), with cantilevers approximately 2, 4, and 8 \um long.  For each device at least two different mechanical modes were detected.  Effective masses ($m_{\rm eff}$) for each mode were computed from dimensions measured with SEM, using FEM to determine the mode shape \cite{Hau13}.  Peaks were thermomechanically calibrated to extract $\omega_0$, the cantilever's resonance frequency, $Q$, the mechanical quality factor in vacuum, and $S_{zz}^{\rm nf}$, the displacement noise floor.  From these parameters we compute $k_{\rm eff}$, the mode's effective spring constant, and \Sfth, the spectral density of thermal forces on the cantilever imposing a force sensing limit.  When measured in air, the quality factors of the cantilevers were reduced by viscous damping and only the out-of plane motion could be detected thermomechanically.  Smaller cantilevers exhibited the larger quality factors in air, and smaller thermal forces, resulting in better force sensing ability - opposite to the case in vacuum. } 
\end{table*}

Despite the slightly larger displacement noise floor, the design presented here has a number of advantages over other optomechanical cantilevers \cite{Li10, Sri11, Liu12, Li12}.  Our devices were fabricated at a commercial foundry (IMEC) using deep UV photolithography with a feature resolution of approximately 130 nm.  This allows for the straightforward, high-throughput fabrication of many such devices for commercial applications, which time-consuming electron beam lithography does not provide.  Further, our devices showcase resonators with two nearly orthogonal modes: in-plane (figure \ref{fig.devs}(d)) and out-of-plane (figure \ref{fig.devs}(e)).  The out-of-plane mode is a better fit to traditional raster scanning or molecule pulling experiments, as current generations of AFMs also operate out-of-plane.  Additionally, these orthogonal modes occur at similar frequencies, providing multi-dimensional force sensing capabilities---a feature especially suited for applications such as protein unfolding experiments, where the unfolding energy landscapes can be highly dependent on the direction of applied force \cite{Bro03}. 

To measure the motion of our device's cantilever,  single mode light from a tunable diode laser (New Focus TLB-6330, 1550-1630 nm) is passed through a dimpled, tapered optical fiber \cite{Mic07} placed on the top edge of the optical microdisk opposite to the mechanical device (figure \ref{fig.how}(b)) using three axes of nanopositioning stages.  By slightly detuning the laser from an optical resonance of the disk, modulations in the frequency of the optical modes induced by the movement of the mechanical resonator are transduced to a voltage signal from a photodetector (PD) measuring the transmission through the tapered fiber.  The high-frequency spectral density of the PD voltage (\Sv) is analyzed to identify peaks corresponding to cantilever modes. A lock-in amplifier (Zurich H2FLI) is then used to measure $\sqrt{S_{VV}}$ across a narrow bandwidth at the mechanical peak frequency, and the polarization and frequency of the tunable laser is iteratively optimized to maximize the mechanical signal (see figure \ref{fig.how}).
Finally, the lock in amplifier is discretely stepped to measure $\sqrt{S_{VV}}$ across the desired frequency range.

By analyzing the power spectral density \cite{Hau13} of laser transmission through the tapered fiber \cite{Mic07} coupled to the microdisk, mechanical motion of the cantilevers can be observed. For each device we were able to identify peaks in the voltage spectral density corresponding to thermodynamic actuation of the fundamental in-plane and out-of-plane modes, while the second out-of-plane mode (figure \ref{fig.devs}(f)) was additionally visible for the 8 \um long cantilever (figure \ref{fig.peaks}(e), (f)). Mode identity was verified by directional piezo actuation and comparison to FEM simulations. The voltage spectral density, \Sv, was calibrated thermomechanically to displacement spectral density (\Sx) of the cantilever's tip. Displacement noise floors of 2 \fmrtHz were observed for the out-of-plane motion of the 4 \um cantilever, equivalent to the best noise floors observed using traditional AFM detection methods \cite{Fukuma2009, Rasool2010}, yet for radically smaller, lighter, and higher-frequency cantilevers.

The linear susceptibility, $\chi(\omega) = z(\omega) / F(\omega)$, relates displacements of the cantilever's tip, $z(\omega)$, to applied forces, $F(\omega)$.  By dividing the measured displacement spectral density by $|\chi(\omega)|^2$, the observed force spectral density can be found (figure \ref{fig.peaks}(b), (d), (f), (h)).  The thermal forces on the cantilever impose a minimum force sensitivity, and in all cases in which the thermomechanical motion of the cantilever was detected, the force noise reached a minimum at the cantilever resonant frequency equal to the thermal noise, ${S}_{FF}^{\rm th} = 4 k_B T m_{\rm eff} \omega_0 / Q$.  For both the in-plane and out-of-plane modes of the 8 \um cantilever, a force sensitivity of $\sim130$ \aNrtHz was achieved, figure \ref{fig.peaks}(b), (d).  This is less than a factor of two higher than a recent hybrid device with a very high mechanical quality factor \cite{Gav12}.  While \Sfth was reached regardless of detector noise, low displacement noise floors broadened the frequency range over which thermally limited force noise was observed.  Therefore, small displacement noise floors, achievable with optomechanics, allow for more accurate, larger bandwidth (faster) force measurements.  A similar effect could be achieved for optomechanical devices such as these through dissipative feedback with optical cooling \cite{Saf12}, broadening the width of the peaks without affecting \Sfth, and allowing wide bandwidth measurements at the thermal noise level for fast scanning \cite{Mertz1993}.

Operating these devices at low bath temperatures would reduce thermal noise on the cantilevers, and a thermal force noise of 1 \aNrtHz at 10 mK is expected to be detectable with the observed noise floors of the 8 \um cantilever's out-of-plane motion, making them excellent candidates for low-temperature precision force measurements \cite{Geraci08}.  Device geometry plays a large part in determining the thermal forces on the cantilevers, and minimizing $m_{\rm eff} \Gamma = m_{\rm eff} \omega_0 / Q$, where $\Gamma$ is the full width half max of the spectral peak, will optimize force sensing ability.  The extremely low effective masses of these devices, ranging here from 140 to 610 fg (table \ref{thetable}), has enabled delicate force sensing despite the modest quality factors of the cantilevers ($\sim \,$ 5000 in vacuum).   

When using the cantilevers to detect forces in liquid or air, thermal force noise on the cantilever is drastically increased due to additional damping of the cantilever, lowering the quality factor of the devices (table \ref{thetable}).  
Higher frequency cantilevers are less affected by viscous dissipation and therefore exhibit better quality factors \cite{Li07, Sun12} and force sensitivity. This is the case for our devices, as the smallest cantilevers are dampened the least in air (see table \ref{thetable}, figure 3(g)-(h), and Appendix D).

\begin{figure}[h]
\centerline{\includegraphics[width=4.5 in]{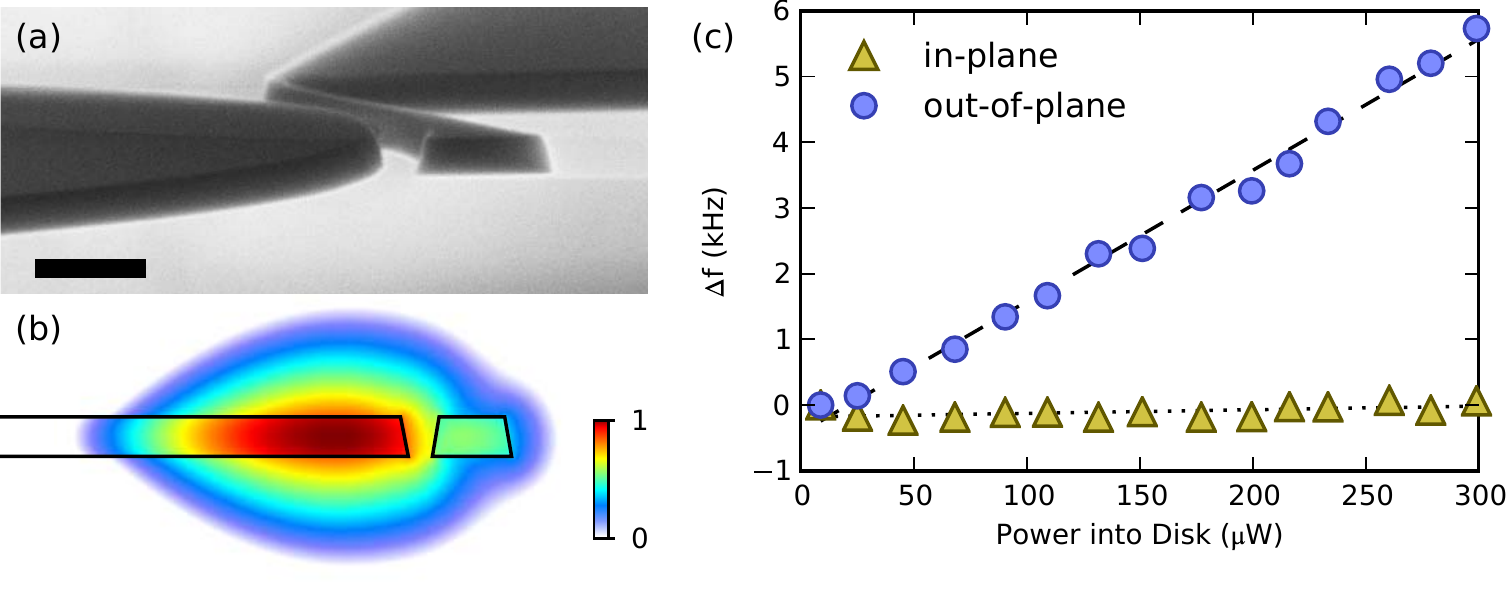}}
\caption{{\label{fig.couple}} (a) Tilted SEM image of a device with a 4 \um cantilever; scale bar 500 nm.  Side walls have a slope of approximately 10$^{\rm o}$ from vertical, creating asymmetries in the optomechanical coupling.  (b)  FEM simulation of an optical mode in cylindrically symmetric coordinates.  Color bar indicates the relative log magnitude of the electric field.  (c)  The whispering gallery cavity modes used to optomechanically detect the cantilever's motion provide an additional restoring force.  By increasing the blue-detuned laser power dropped into the optical disk used to detect the 4 \um cantilever's motion, stiffening of the cantilever is observed.  The frequency of the out-of-plane motion increased by $\sim\,0.1$\%, while the in-plane motion showed negligible effect due to its $\sim 100 \times$ smaller $g_{\rm om}$, table \ref{thetable}.  Error bars from numerical fits are smaller than the marker size, and dashed lines are guides to the eye.
}
\end{figure}

The small displacement noise floors achieved with these devices are a direct result of the efficiency in which displacements of the cantilever are transduced into frequency changes in the optical disk.  This efficiency can be described to first order by the optomechanical coupling coefficient, $g_{\rm om} = d\omega_{\rm opt} / dz$, where $\omega_{\rm opt}$ is the optical mode resonance frequency.  Cantilevers curve with the microdisk to optimize $g_{\rm om}$ by increasing overlap between the optical whispering gallery modes and the cantilever's motion (table \ref{thetable}).  In all devices, the out-of-plane motion of the cantilever had considerably better optomechanical coupling than the in-plane motion.  The apparent symmetry of the out-of-plane motion would suggest a very small linear optomechanical coupling for the out-of-plane mode, however slanted sidewalls of the devices due to fabrication (figure \ref{fig.couple}(a),(b)), and the placement of the dimpled fiber touching the top of the optical disk introduce sufficient asymmetries to explain the large linear optomechanical coupling observed \cite{Kim13}.

Finally, optomechanics provides schemes to introduce feedback and control over AFM cantilevers, such as optomechanical heating and cooling of the mechanical modes, and optical gradient forces.  For example, we show in figure 4(c) the optical spring effect can be used to tune the cantilever frequency \textit{in situ} by controlling the optical gradient force through the laser power used for detection.  Such techniques introduce new methods for manipulating optomechanical AFMs for increased functionality.

\section{Conclusion}

Optomechanical AFM provides the path to ultra-sensitive molecular force probe spectroscopy, HS-AFM, and other AFM applications.  We have demonstrated optomechanical detection of sub-picogram effective mass multidimensional AFM cantilevers that are commercially fabricated, with displacement noise floors down to 2 fm\rtHz, and 130 aN\rtHz force sensitivity at room temperature.  Challenges remain, including selective attachment of relevant molecules, yet we envision that extension of the devices presented here to aqueous environments will open new doors in high-speed, high-resolution molecular force measurements.

\ack

The authors wish to thank University of Alberta, Faculty of Science; the Canada Foundation for Innovation; the Natural Sciences and Engineering Research Council, Canada; and Alberta Innovates Technology Futures for their generous support of this research.  We also thank Greg Popowich for  technical support, along with Mark Freeman, Wayne Hiebert and Paul Barclay for helpful discussions.

\appendix

\section{Thermomechanical Calibration}

To calibrate the voltage spectral density, $S_{VV}$, measured from a photodetector into a displacement spectral density, $S_{zz}$, of the movement of the cantilever's tip, the thermal forces on the cantilever can be used.
By way of the fluctuation-dissipation theorem, the thermal forces acting on a cantilever's mode are constant across frequencies with a spectral density of \cite{Gav12, Hau13}

\begin{equation}
S_{FF}^{\rm th} = 4 k_B T m_{\rm eff} \omega_0 / Q,
\end{equation}
where $k_B$ is the Boltzmann constant, $T$ is the system temperature, $\omega_0$ is the mode's resonance frequency, $Q$ is the quality factor, and $m_{\rm eff}$ is the effective mass of the cantilever (described further below).
Using the linear susceptibility of a damped harmonic oscillator, 

\begin{equation}
\chi(\omega) = \frac{1}{m_{\rm eff} [\omega_0^2 - \omega^2 - i\omega \omega_0/Q]},
\end{equation}
the theoretical displacement spectral density corresponding to thermomechanical actuation of the cantilever mode is known: $S_{zz}^{\rm th}(\omega) = |\chi(\omega)|^2 S_{FF}^{\rm th}$.  Further, assuming the voltage measured is linearly proportional to cantilever displacement, and the noise from the measurement apparatus is constant across frequencies of interest, a theoretical fit to the voltage spectrum can be found \cite{Bunch07},

\begin{equation}
S_{VV}(\omega) = S_{VV}^{\rm nf} + \alpha S_{zz}^{\rm th}(\omega),
\end{equation}
where $S_{VV}^{\rm nf}$ is the voltage noise floor density and  $\alpha$ is a conversion factor between volts and meters, ie. $\alpha = (\mathrm{d}V / \mathrm{d}z)^2$.  Substituting in the thermal displacement noise,

\begin{equation}\label{eq.fit}
S_{VV}(\omega) = S_{VV}^{\rm nf} + \alpha \frac{ 4 k_B T \omega_0}{m_{\rm eff} Q} \frac{1}{(\omega_0^2 - \omega^2)^2  + (\omega \omega_0 / Q)^2}.
\end{equation}
Because it is not possible to differentiate both $\alpha$ and $m_{\rm eff}$ from the fit,  $m_{\rm eff}$ is calculated beforehand from measured cantilever dimensions.  By modelling the structural modes of the cantilever using the finite element method (FEM), the mode shape of interest, $\mathbf{r}(\mathbf{x})$, which is the mechanical displacement of the mode from it's undeformed position, $\mathbf{x}$, normalized to the maximum displacement, can be determined and the effective mass can be computed by carrying out an integral over the volume of the cantilever \cite{Eichenfield2009, Hau13},

\begin{equation}
m_{\rm eff} = \int \mathrm{d}V \rho(\mathbf{x}) |\mathbf{r}(\mathbf{x})|^2.
\end{equation}
By fitting the measured $S_{VV}$ to (\ref{eq.fit}),  the resonance frequency ($\omega_0$), quality factor ($Q$), noise floor ($S_{VV}^{\rm nf}$), and the voltage-displacement conversion factor ($\alpha$) used to calibrate the spectrum, can be determined.  Calibrated displacement spectral densities and force spectral densities for the cantilevers discussed above are shown in Appendix D.

\section{Determining the Optomechanical Coupling Coefficient}

By performing thermomechanical calibration, the voltage-displacement conversion factor, $\alpha$, was found.  Since $\alpha$ linearly converts displacements of the cantilever ($S_{zz}$) to volts from the photodetector ($S_{VV}$), $\sqrt{\alpha} = \mathrm{d}V / \mathrm{d}z$.  Examining the optomechanical detection mechanism, the displacement to voltage transduction can be divided into two steps,  displacement to optical cavity frequency ($\omega_{\rm opt}$) shifts, and $\omega_{\rm opt}$ to transmission (voltage) transduction.  Therefore, with help of the chain rule, $\mathrm{d}V / \mathrm{d}z = (\mathrm{d}V / \mathrm{d}\omega_{\rm opt}) (\mathrm{d}\omega_{\rm opt} / \mathrm{d}z$).  Here $\mathrm{d}\omega_{\rm opt} / \mathrm{d}z$ is the optomechanical coupling coefficient, $g_{\rm om}$ \cite{Eichenfield2009, Anetsberger2009}.  By calculating the slope of laser transmission \textit{vs.} laser frequency at the frequency of light the mechanical signal was detected at (\textit{e.g.} from figure 2(c) in the main text), $\mathrm{d}V / \mathrm{d}\omega_{\rm opt}$ can be determined, enabling calculation of $g_{\rm om}$.

\section{Optical Power Calibration}

The optical power coupled into the microdisk was estimated by splitting off a small portion (10\%) of the light before the vacuum chamber and sending it to a power meter (Thorlabs PM100D), instead of the wavelength meter (figure 2(a)). Power meter readings were calibrated to photodector voltages in the absence of an optical resonator to compensate for the wavelength-dependent response of the photodetector and intrinsic resonances in the fibers used. The total optical power in the fiber was then calculated by monitoring the power at the small split-off as the attenuation was modified using the variable attenuator. The net power dropped into the disk was found by comparing the transmission through the dimpled fiber before and after coupling to the optomechanical devices.

\section{Additional Data}

\begin{figure}[H]
\centerline{\includegraphics[width=6.0in]{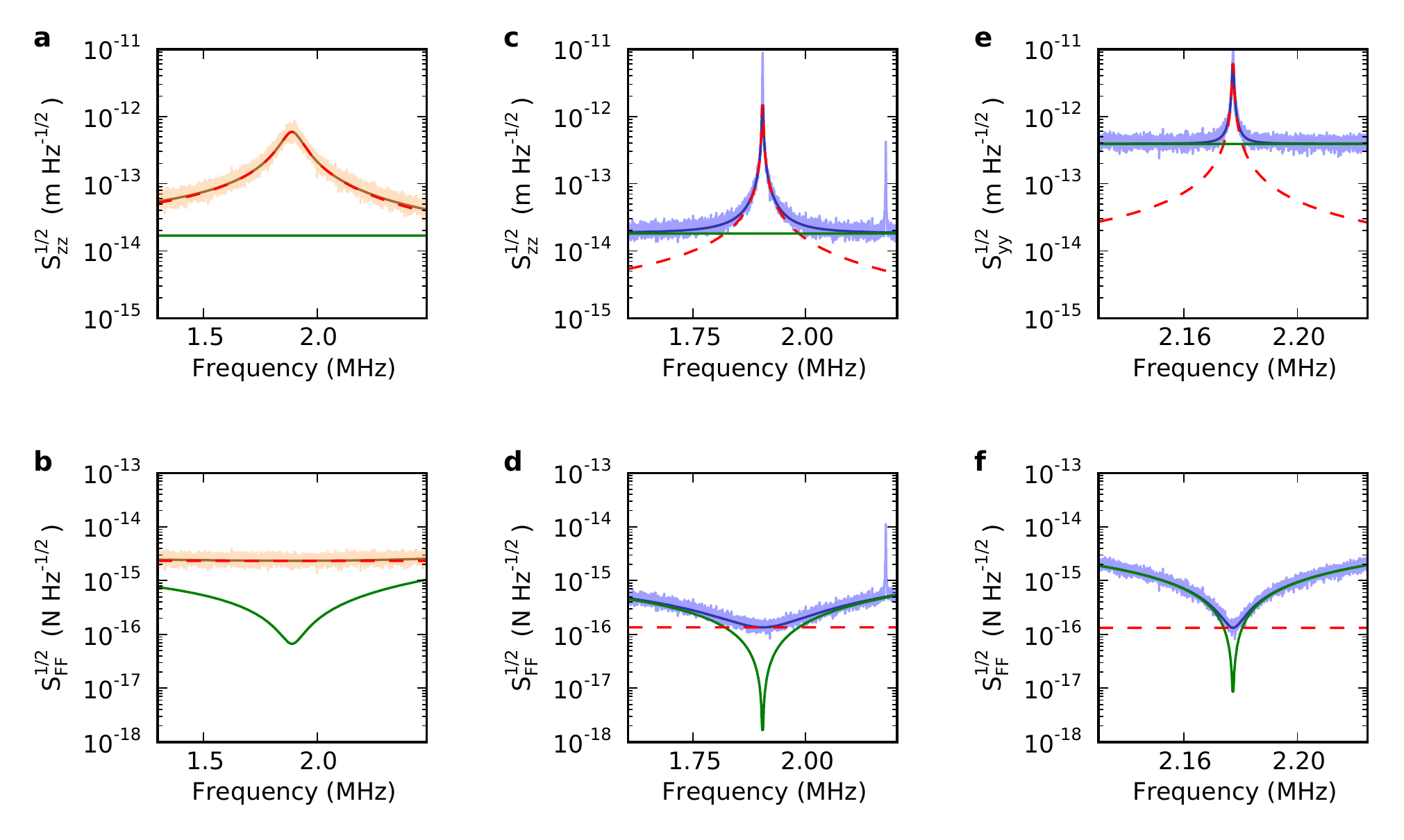}}
\caption{{\label{fig.peaks8}} \textbf{Spectral peaks for the 8 $\mathbf{\boldsymbol\mu m}$ cantilever.} \textbf{a},  Spectral displacement density and \textbf{b}, spectral force density corresponding to thermal motion of the out-of-plane mode at atmospheric pressure. \textbf{c},  \textbf{d}  out-of plane motion in vacuum, and \textbf{e}, \textbf{f}, in-plane motion in vacuum.
}
\end{figure}

\begin{figure}[H]
\centerline{\includegraphics[width=6.0in]{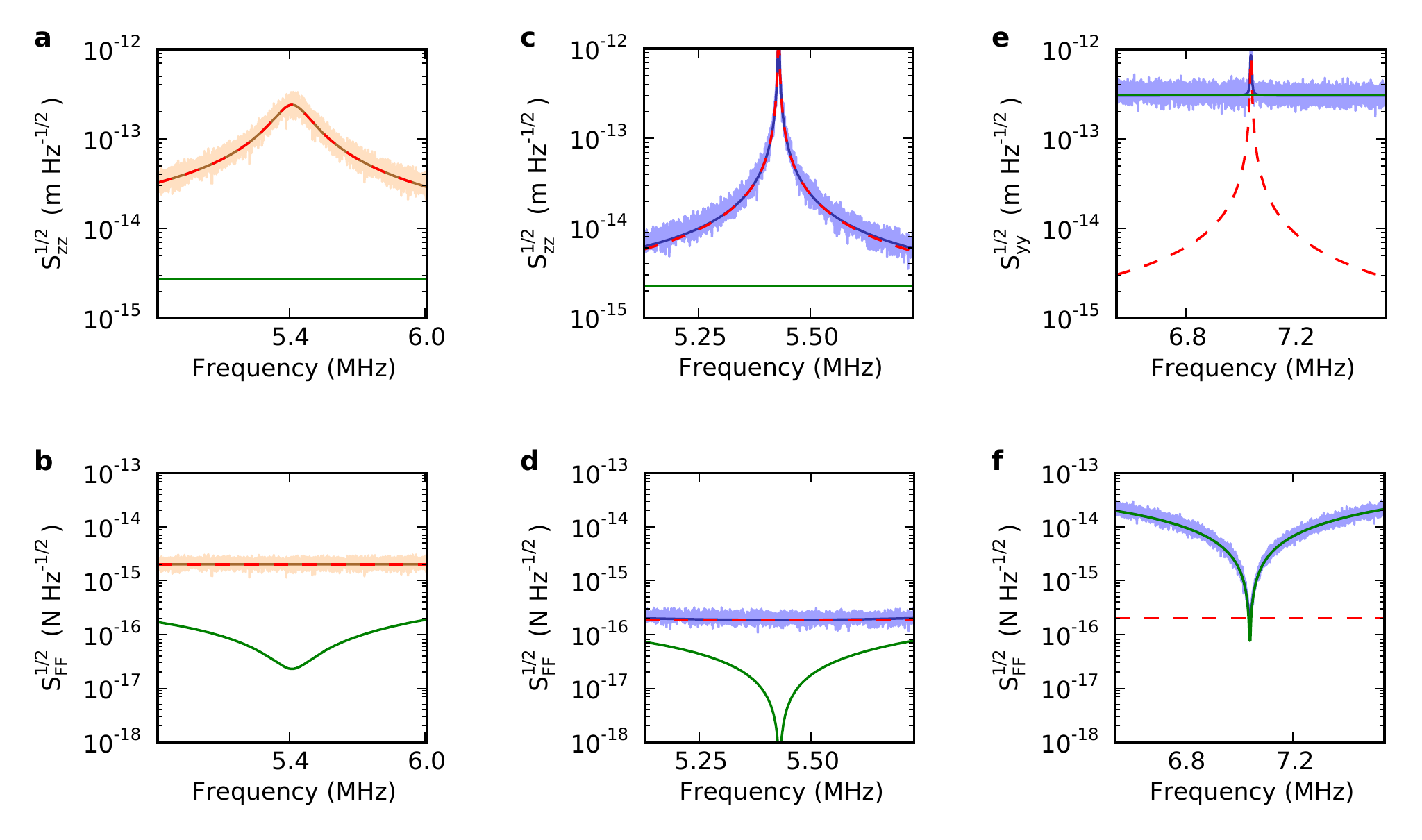}}
\caption{{\label{fig.peaks4}} \textbf{Spectral peaks for the 4 $\mathbf{\boldsymbol\mu m}$ cantilever.} \textbf{a},  Spectral displacement density and \textbf{b}, spectral force density corresponding to thermal motion of the out-of-plane mode at atmospheric pressure. \textbf{c},  \textbf{d}  out-of plane motion in vacuum, and \textbf{e}, \textbf{f}, in-plane motion in vacuum.
}
\end{figure}

\begin{figure}[H]
\centerline{\includegraphics[width=6.0in]{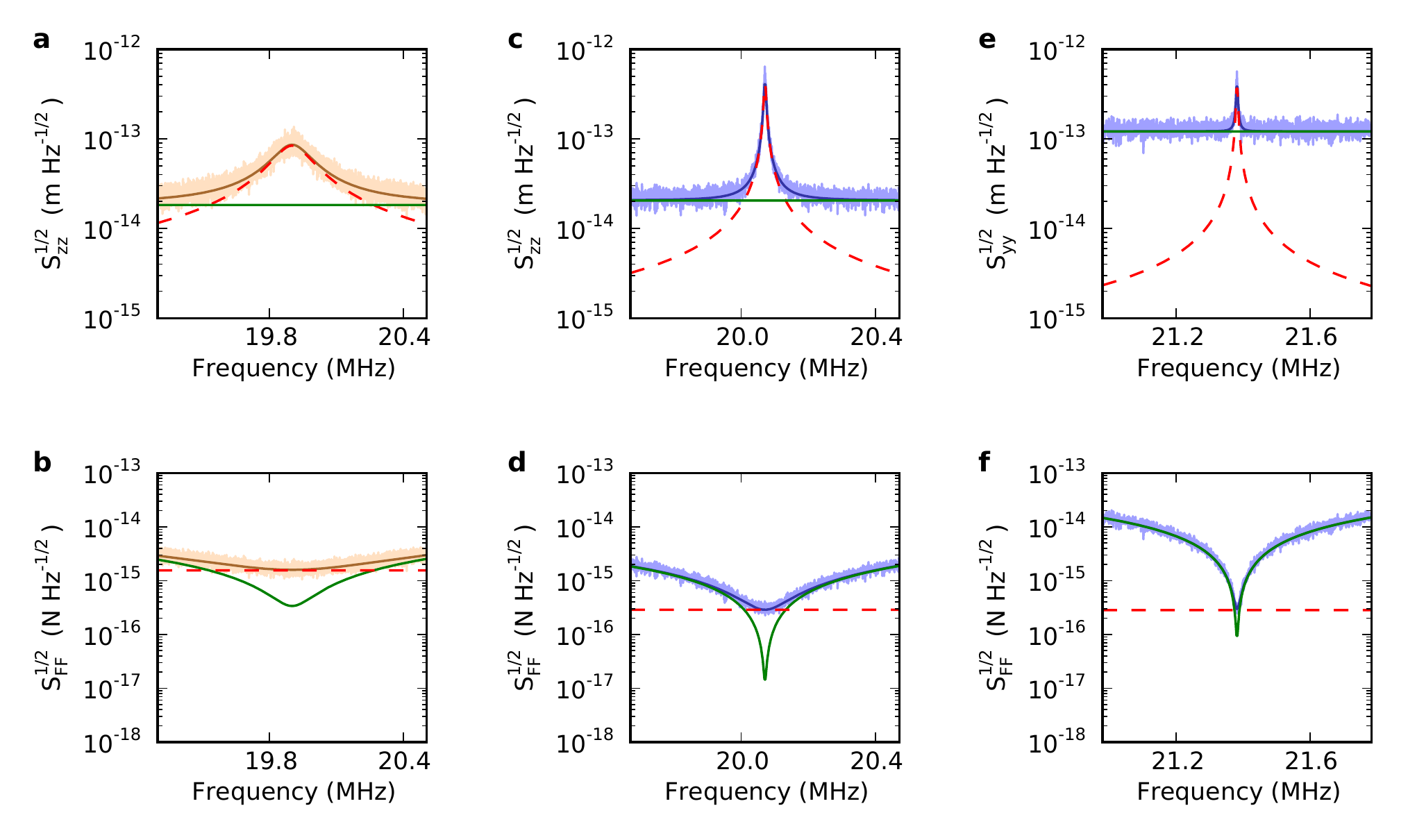}}
\caption{{\label{fig.peaks2}} \textbf{Spectral peaks for the 2 $\mathbf{\boldsymbol\mu m}$ cantilever.} \textbf{a},  Spectral displacement density and \textbf{b}, spectral force density corresponding to thermal motion of the out-of-plane mode at atmospheric pressure. \textbf{c},  \textbf{d}  out-of plane motion in vacuum, and \textbf{e}, \textbf{f}, in-plane motion in vacuum.
}
\end{figure}

\end{document}